\documentclass[twocolumn]{revtex4}
\usepackage{amsmath} \usepackage{graphicx} \usepackage{dcolumn}
\usepackage{bm} \usepackage{epsf} \usepackage{amssymb}

\begin{document}

\newlength{\figurewidth}
\setlength{\figurewidth}{\columnwidth}

\sloppy

\newcommand{\prtl}{\partial}
\newcommand{\la}{\left\langle}
\newcommand{\ra}{\right\rangle}
\newcommand{\dla}{\la \! \! \! \la}
\newcommand{\dra}{\ra \! \! \! \ra}
\newcommand{\we}{\widetilde}
\newcommand{\smfp}{{\mbox{\scriptsize mfp}}}
\newcommand{\sLZ}{{\mbox{\scriptsize LZ}}}
\newcommand{\sph}{{\mbox{\scriptsize ph}}}
\newcommand{\sinhom}{{\mbox{\scriptsize inhom}}}
\newcommand{\sneigh}{{\mbox{\scriptsize neigh}}}
\newcommand{\srlxn}{{\mbox{\scriptsize rlxn}}}
\newcommand{\svibr}{{\mbox{\scriptsize vibr}}}
\newcommand{\smicro}{{\mbox{\scriptsize micro}}}
\newcommand{\smax}{{\mbox{\scriptsize max}}}
\newcommand{\seq}{{\mbox{\scriptsize eq}}}
\newcommand{\sstr}{{\mbox{\scriptsize str}}}
\newcommand{\teq}{{\mbox{\tiny eq}}}
\newcommand{\sinn}{{\mbox{\scriptsize in}}}
\newcommand{\tin}{{\mbox{\tiny in}}}
\newcommand{\scr}{{\mbox{\scriptsize cr}}}
\newcommand{\sscr}{{\mbox{\scriptsize scr}}}
\newcommand{\sL}{{\mbox{\scriptsize L}}}
\newcommand{\sTS}{{\mbox{\scriptsize TS}}}
\newcommand{\stheor}{{\mbox{\scriptsize theor}}}
\newcommand{\sGS}{{\mbox{\scriptsize GS}}}
\newcommand{\sNMT}{{\mbox{\scriptsize NMT}}}
\newcommand{\sRFOT}{{\mbox{\scriptsize RFOT}}}
\newcommand{\tRFOT}{{\mbox{\tiny RFOT}}}
\newcommand{\sbulk}{{\mbox{\scriptsize bulk}}}
\newcommand{\tbulk}{{\mbox{\tiny bulk}}}
\newcommand{\sauto}{{\mbox{\scriptsize auto}}}
\newcommand{\tauto}{{\mbox{\tiny auto}}}
\newcommand{\sescape}{{\mbox{\scriptsize escape}}}
\newcommand{\sth}{{\mbox{\scriptsize th}}}
\newcommand{\svib}{{\mbox{\scriptsize vib}}}
\def\Xint#1{\mathchoice
   {\XXint\displaystyle\textstyle{#1}}%
   {\XXint\textstyle\scriptstyle{#1}}%
   {\XXint\scriptstyle\scriptscriptstyle{#1}}%
   {\XXint\scriptscriptstyle\scriptscriptstyle{#1}}%
   \!\int}
\def\XXint#1#2#3{{\setbox0=\hbox{$#1{#2#3}{\int}$}
     \vcenter{\hbox{$#2#3$}}\kern-.5\wd0}}
\def\ddashint{\Xint=}
\def\dashint{\Xint-}
\title{A Universal Criterion of Melting}

\author{Vassiliy Lubchenko} \affiliation{Department of Chemistry,
  University of Houston, Houston, TX 77204-5003}


\begin{abstract}

  Melting is analyzed dynamically as a problem of localization at a
  liquid-solid interface. A Lindemann-like criterion of melting is
  derived in terms of particular vibrational amplitudes, which turn
  out to equal a universal quotient (about one-tenth) of the molecular
  spacing, at the interface. The near universality of the Lindemann
  ratio apparently arises owing to strongly overdamped dynamics near
  melting, and despite the anharmonic interactions' being
  system-specific. A similar criterion is derived for structural
  displacements in the bulk of the {\em solid}, in particular the
  premelted layer; the criterion is no longer strictly universal, but
  still depends only on the harmonic properties of the solid. We
  further compute the dependence of the magnitude of the elemental
  molecular translations, in deeply supercooled fluids, on the
  temperature and the high frequency elastic constants.  We show
  explicitly that the surface tension between distinct liquid states,
  near the glass transition of a supercooled liquid, is nearly evenly
  split between entropic and energetic contributions.

\end{abstract}

\date{\today}

\maketitle

\section{Introduction: what is ``melting''?}

The word ``melting'' turns out to mean more than one thing. (For
reviews, see \cite{Dash_ContPhys, Dash_RMP, Sorkin_MSthesis, Sorkin,
  Ubbelohde, Roberts, BornHuang, Bilgram, Nabarro, Bienfait,
  SurfRough, Rosenberg}.)

At a high enough temperature, the shear modulus of a perfect crystal
would vanish, which, according to Born \cite{BornHuang, Yip_stab},
would lead to melting.  This type of melting is sometimes called {\em
  mechanical} melting.  Experience shows nevertheless that in real
crystals, the lattice always begins to desintegrate at significantly
lower temperatures, when the shear resistance of the bulk still
exceeds at least a half of its zero temperature value
\cite{Dash_ContPhys, Dash_RMP, Sorkin_MSthesis}.  This type of
melting, called {\em thermodynamic} melting, appears to usually
initiate at defects, most notably at the free surface, but also at
grain boundaries, impurity sites etc. \cite{Dash_ContPhys, Dash_RMP}.
Another well known peculiarity of the melting/fusion transition is,
while liquids are relatively easy to overcool, overheating a crystal
seems very difficult, except under special circumstances, such as when
the free surface is ``clamped'', or in the case of water, which
contracts upon melting thus allowing for internal melting
\cite{Bilgram}.  Surface melting phenomena are very complicated, as is
just about anything about surfaces: The melt may or may not wet the
crystal surface; the thickness of the ``premelted'' layer, separating
the crystal from the vapor, often exhibits power law scaling with the
proximity to the melting temperature; the surface itself is usually
reconstructed and often experiences roughening transitions below
melting \cite{Bienfait, SurfRough}. One should add here the effects of
lattice anisotropy \cite{Frenken, Lipowsky} and possible polymorphic
transitions near the melting temperature \cite{GoetzHergenrother}. To
summarize, conceptualizing melting as a dynamical process is not
straightforward and depends on specific circumstances.

In view of the complications above, it comes as a surprise that there
should be the following simple, nearly universal phenomenological
criterion, due to Lindemann \cite{Lindemann, Roberts, Frenkel}: At
melting the typical vibrational displacement, for a given crystalline
class, should be some {\em fixed} fraction of the lattice
spacing. Lindemann posited that the ratio should be about a half,
implying direct {\em collisions} between the atoms constituting the
lattice would become possible, leading to the lattice's demise. One
may note that the Lindemann's argument accounts for anharmonicities in
the problem but in a very generic fashion, through the {\em existence}
of collisions.  The value of the Lindemann ratio was later revised by
Gilvarry \cite{Gilvarry} to be about one tenth and works rather well
indeed: Data compilations \cite{Bilgram, GuptaSharma} show only a
variation of 10\% or so within a given crystal structure type, and the
overall range is between 0.068 and 0.114.

It is in the context of the Einstein's theory of vibrations in solids
that Lindemann formulated his criterion; perhaps for this reason, the
criterion has been a benchmark in density functional studies (DFT) of
crystal melting (see e.g. \cite{Baus, Haymet}), which are accurate in
the short-wavelength, Einstein limit.  Nevertheless, it was not until
a DFT study of {\em aperiodic} crystals, when the Lindemann ratio
turned up on a {\em first principles} basis, as an order parameter:
Wolynes and coworkers \cite{dens_F1, dens_F2} demonstrated that a
liquid, if failed to crystallize, should settle into (long-living)
aperiodic structures.  The transition, or rather a cross-over, is
characterized by a discontinuous change in the localization length
from an effectively infinite to a finite value; the latter gives the
vibrational displacement at the mechanical stability edge. This length
therefore directly corresponds to the Lindemann length; its
DFT-computed value matches well that observed in crystals, and, in the
first place, the neutron scattering data in supercooled liquids and
frozen glasses \cite{Mezei, MezeiRussina}.  Lindemann-like criteria
also naturally arise in treatments of energy landscape models of
protein folding and collapse \cite{SasaiWolynes, SasaiWolynes_PRA},
mean-field models of the structural glass transition \cite{repl_Lind},
and in vortex lattice melting in superconductors \cite{Hwa}.  Even in
the absence of a first principles justification, the Lindemann rule is
often used on purely empirical grounds, evidently owing to its
remarkable circumstantial consistency, and its simplicity.
Applications range from vortex lattices in rotating Bose condensates
\cite{Baym} to estimating the native state entropy of a protein
\cite{LWF}. {\em Generalized} Lindemann criteria have been applied to
defect-induced amorphization of a crystal \cite{Okamoto}, or melting
in one-component plasmas (see \cite{Ceperley_OCP} and references
therein). An inverse Lindemann criterion has been suggested for
crystallization \cite{Stillinger_InvLind}.  An increasingly useful
application of the Lindemann criterion is in molecular dynamics
simulations (see e.g. \cite{Karplus_Lindemann}), where definitive
observation of melting is usually beyond current computational
technology.

Perhaps, the most immediate objection to the Lindemann's criterion is
that it involves characteristics of only {\em one} of the two phases
coexisting at melting; a proper criterion, presumably, should compare
some property of {\em both} phases. For example, in the absence of
extensive defects, comparing the bulk free energies would be
adequate. In addition, as we now understand, a proper analysis of
dynamical melting should proceed with reference to processes at the
liquid/solid interface. The present work implements these two notions
in the following way:

First, in Subsection \ref{Interface}, we consider the escape of a
molecule from the solid/liquid interface into the liquid.  Here, two
length scales of molecular motions will arise, whose ratio to the
molecular spacing is universal at melting.  One length scale is the
size of the metastable minimum harboring a molecule which is about to
change its location (for example, to exit into the liquid); the other
is the extent of the {\em transition state} during the exit. The
relation of the two scales to the {\em vibrational amplitudes} proper
will depend on the detailed morphology of the region in question; so
for example, the surface roughness or the specific crystal face will
affect surface melting. Under most circumstances, nevertheless, the
derived criterion will simply amount to a simple Lindemann-like
criterion.

Second, we ask in Subsection \ref{L}, what would be the magnitude of
the (weakly activated) molecular displacements in the {\em bulk of the
  solid}, in the presence of alternative structural states. For
glass-forming substances, these displacements are actually present in
the bulk of the material, if it is supercooled. Otherwise, alternative
structural states are present only in a ``premelted'' layer, if any.
A proper melting criterion will be formulated, which amounts to {\em
  comparing} the vibrational molecular amplitudes to the structural
displacements: in a stable solid, the former should less than the
latter. We will further deduce the dependence of the structural
displacements on the material's stiffness, and the temperature.  The
results apply directly to premelted layers and supercooled liquids.

\section{Derivation of a melting criterion}

\subsection{Melting at the Interface}
\label{Interface}

It will be most straightforward to see how a melting criterion arises
for molecules that are directly at the solid-liquid interface, in the
sense that here, simply an {\em isolated} activated event is required
in order for a molecule to exit irreversibly into the liquid.  A
number of conventions as to what an ``interface'' is are possible and
are subject to the same ambiguities as the definitions of the phases
themselves.  It is beyond reasonable doubt that the surface region of
a melting solid is far from the simplified text-book pictures, even
ignoring surface roughening effects: The interface is not sharp, and
often extends for up to several tens of atoms, as could be deduced,
somewhat indirectly, from studies of (pre)melted layers demarcating a
solid from its vapor \cite{Frenken, ZhuDash}. (See also a recent
review \cite{Rosenberg} on surface ice melting.)  That the interface
is ``diffuse'' near melting has been also concluded theoretically, via
density functional studies \cite{OxtobyHaymet} employing specific
ansatzes for the equilibrium distribution functions in the solid and
liquid.  Furthermore, the heterogeneity across the layer is not only
structural but must also involve a heterogeneity in relaxation times,
i.e. the lifetimes of long-living local structures.  Similar to the
density, the life-times in the premelt layer interpolate between those
in the liquid and the solid.  Several ways to deal with the ambiguity
in defining a solid-liquid interface may be proposed. For example,
Oxtoby and Haymet \cite{OxtobyHaymet} employ an appropriate
equilibrium order parameter changing continuously when going from
liquid to solid.  Trayanov and Tosatti \cite{TrayanovTosatti} analyze
premelted layers, in a mean-filed fashion, based on two order
parameters, ``density'' and ``crystallinity''.  Here, in order to
focus on the dynamical aspects of melting, we will use the life-times
of local metastable structures to establish an operational criterion
of whether a molecule is in the solid, or part of the liquid.  While
the equlibrium interfacial region may be discussed only in a broad
sense, as a ``diffuse'' entity, the dynamic interface will turn out to
be thin and well defined.

We will distinguish between the liquid and the solid in the usual way,
via symmetry, and will specifically focus on the time scale on which
the symmetry is broken/restored.  Consider a substance consisting of a
single, relatively compact molecular species, in the classical
regime. The crystal breaks the translational symmetry in that here,
one can {\em label} the molecules based solely on their each being
located within a particular, well defined cell. (C.f. however the
incommensurate quantum crystals \cite{PWA_incomm}).  One may speak of
a fluid, on the other hand, when such labelling is impossible. The
corresponding translational symmetry is physically maintained by
particle transport.  Call $\tau_0$ the time it takes a molecule to
diffuse a distance defining the volumetric density of the
liquid. Choose a compact, specific cell, in space, whose volume is
equal to the volume per molecule in the liquid.  Since the time
$\tau_0$ is significantly longer than the time scale of density
fluctuations, it is guaranteed that {\em another}, identical molecule
will have visited the chosen cell, within time $\tau_0$ upon the exit
of the previous cell's dweller, thus erasing the possibility that one
be able to label a molecule by its spatial location. It is therefore
at times exceeding $\tau_0$ that one may speak of a liquid
state. Recall also that we are considering compact molecules, and so
rotational diffusion/equilibration is not an issue.  It is useful, for
future reference, to compute the particle exchange time $\tau_0$, in
terms of the collisional time: Let $1/a^3 \equiv n$ be the molecular
concentration in the fluid, so that $a$ is the average, {\em
  volumetric} molecular spacing. The typical collisional, or
auto-correlation time (also defining the density fluctuation
time-scale) is $\tau_{\sauto} = m/\zeta$, where $\zeta \simeq 6 \pi a
\eta$ is the friction coefficient, $\eta$ is the viscosity, and $m$ is
the molecule's mass. The time $\tau_0$ it takes to diffuse a distance
$a$ is roughly $a^2/6D$, where $D$ is the diffusion constant, related
to the friction through the Einstein's relation $D = k_B T/\zeta$. As
a result,
\begin{equation} \label{t_t}
  \frac{\tau_0}{\tau_{\sauto}} \simeq \frac{6 \pi^2 \eta^2 a}{\rho k_B T},
\end{equation}
where $\rho$ is the liquid's mass density.  One may directly check
that near melting, this ratio is generically about $10^3$ but varies
within an order of magnitude or so between different substances: For
instance, cobalt and sodium yield $1.3 \cdot 10^3$ and $2.1 \cdot
10^3$ respectively.  This indicates it takes about a thousand
molecular collisions or so, per molecular volume, to establish local
thermal equilibrium in a liquid. Finally, note that the large value of
the ratio in Eq.(\ref{t_t}) is an internal test of the argument's
consistency: To give a counter example, it would be incorrect to use
analogous logic to estimate equilibration times in {\em dilute}
gases. Here one would find, using the elementary kinetic theory, that
$\tau_0/\tau_\sauto \simeq 9 (\sigma/a^2)^2 < 1$, where $\sigma$ is
the molecular scattering cross section; clearly $\tau_0$ does not
correspond to an equilibration time scale.  Of course, the rate
limiting step during equilibration in dilute gases is diffusion in the
{\em momentum} space, which is responsible for establishing the
Maxwell distribution of velocities; whereas in dense liquids near
fusion, the rate limiting step is configurational equilibration.

Consider now a region of space occupied by a solid and its melt, at
some temperature $T$ just above the lowest temperature, $T_m$, at
which surface melting is possible.  Suppose there is a molecule, in
the region, that fails to move a distance $a$ in the time
$\tau_0$. (To avoid confusion we note that if long wave-length sound
is present in the system, one should stipulate that the local
reference frame move with the sound.)  All such molecules can not be
regarded as part of the liquid, as just discussed, and so we must
regard them as part of the solid. The boundary of any (spatially)
closed set of such molecules may therefore be defined as the
solid-liquid {\em interface}. On the other hand, the inability of a
molecule to move the distance $a$ in time $\tau_0$ is equivalent to
saying the molecule is residing in a metastable free energy
minimum. In other words, a molecule is part of the solid, if the
escape time $\tau_\sescape$ from its current neighborhood exceeds the
exchange time $\tau_0$:
\begin{equation}
  \label{t>t}
\tau_\sescape > \tau_0.
\end{equation} 
The escape time $\tau_\sescape$ will generally differ for distinct
crystalline faces, or various distinct surface morphologies. The above
stipulation that
\begin{equation}
  T = T_m + 0^+,
\end{equation}
implies that there is at least one specific face/morphology which is
in near equilibrium with the liquid, and there are no
face/morphologies which are melting in a spontaneous fashion. In the
following we will specifically consider those faces that are melting
in a quasi-equilibrium fashion. For these,
\begin{equation} \label{t=t}
  \tau_\sescape = \tau_0.
\end{equation}
One may regard this expression as the dynamical definition of a
solid/liquid interface.  A melting temperature $T_m$, as defined
above, will generally differ from the usual calorimetric melting
temperature, if the solid melts anisotropically. (The latter is
usually the case.)

\begin{figure}[t]
\includegraphics[width=.75\figurewidth]{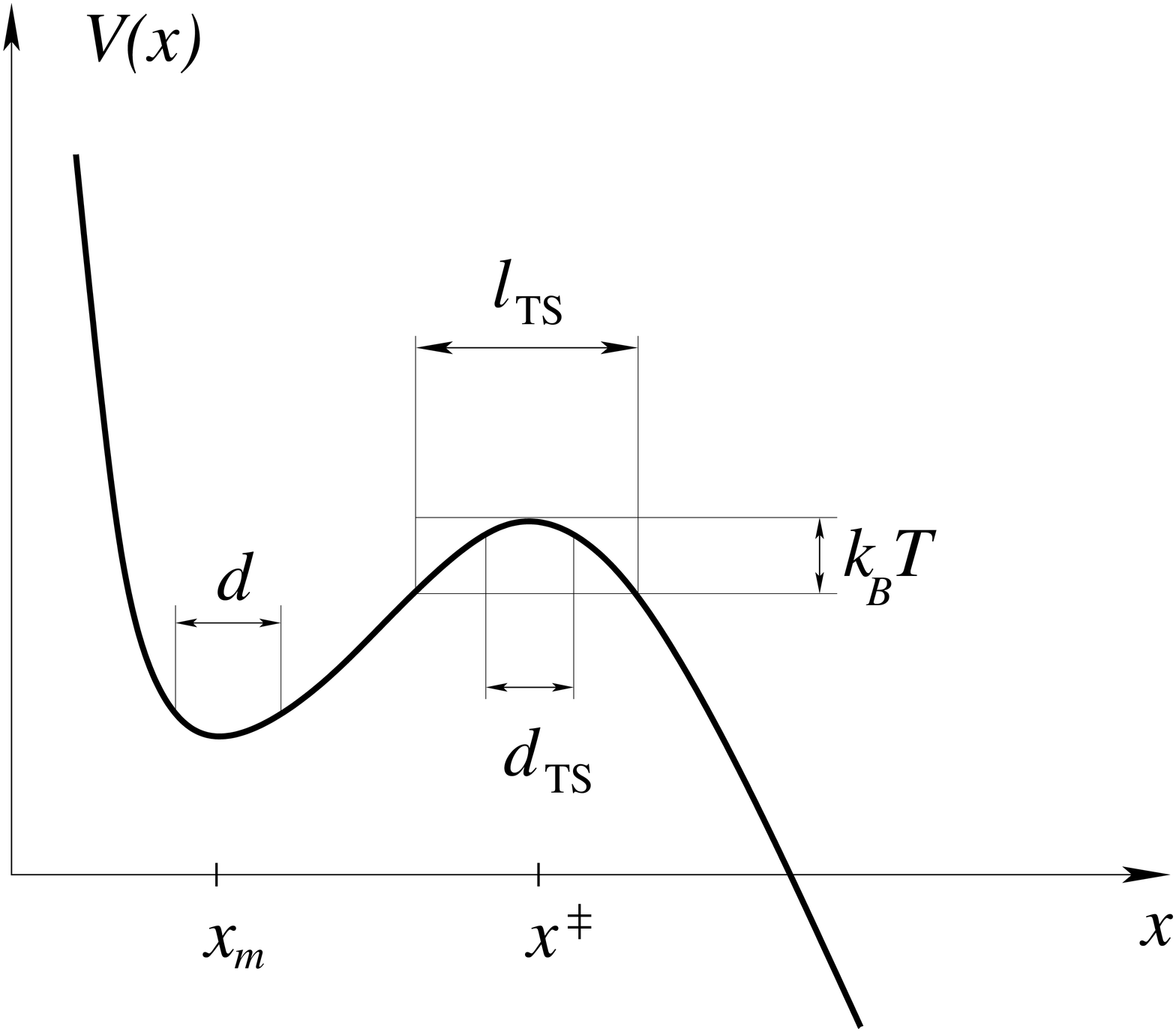}
\caption{\label{rxn_sch} A generic schematic of an escape free energy
  profile is shown. $l_\sTS$, the transition state size, demarcates
  the vicinity of the saddle point within the thermal energy from the
  top. $d$ and $d_\sTS$ are defined in text.}
\end{figure}

In order to estimate $\tau_\sescape$, we shall adopt the approach of
Frauenfelder and Wolynes (FW) \cite{FW} (see also \cite{Kramers,
  Hanggi_RMP}), who have delineated the various activated transport
regimes, and computed the corresponding rates, in terms of several
characteristic {\em length} scales.  First note that owing to the
frequent collisions (see also below), the motion of the reaction
coordinate corresponding to the escape mode, is strongly overdamped.
An operational criterion of this is that the particle's mean free path
be significantly shorter than the transition state size: $l_\smfp \ll
l_\sTS$. (A schematic of the free energy profile along the progress
coordinate is shown in Fig.\ref{rxn_sch}.) As a result, the particle
stays in the transition state region for a long time, relative to the
molecular collisional time $\tau_{\sauto}$, leading to a large number
of barrier crossings, while at the top of the barrier: $N_c =
l_\sTS/l_\smfp$.  The corresponding rate may therefore be estimated
using the standard transition state result, multiplied by a (small)
transmition factor $\kappa \simeq 2 l_\smfp/l_\sTS = 2 \tau_\sauto
v_{\sth}/l_\sTS$, as appropriate in the overdamped Kramers
limit. Here, $v_\sth = \sqrt{3 k_b T/m}$ is the thermal velocity of
the particle.
As a result, the Kramers rate reads:
\begin{equation} \label{tauescape} \tau_\sescape^{-1} = \frac{1}{2}
  \frac{\la |v| \ra}{(\sqrt{2 \pi} d)} \frac{2 \tau_\sauto
    v_\sth}{l_\sTS} e^{-V^\ddagger/k_B T},
\end{equation}
where
\begin{equation} \label{d}
  d \equiv \frac{1}{\sqrt{2 \pi}} \int_{x_m} dx \exp^{-[V(x) - V(x_m)]/k_B T}
\end{equation}
is the ``size'' of the metastable basin. (The label $x_m$ in the
integration limits indicates that the integral is understood in its
asymptotic sense in terms of the basin's width.)  The coefficient
$1/\sqrt{2 \pi}$ was incorporated in the definition so that $d = \la
(\Delta x)^2 \ra^{1/2}$, if the potential is strictly harmonic at the
minimum $x_m$. $\la | v | \ra = \sqrt{2/(3 \pi)} v_\sth$ is the
thermally averaged particle speed that enters the expression for the
molecular flux at the barrier top. Lastly, $V^\ddagger \equiv
V(x^\ddagger) - V(x_m)$ is the barrier height. The latter is actually
quite easy to evaluate after one realizes that an escape event will
have occured, if a particle's displacement just exceeds the typical
thermal vibrational amplitude {\em of} the interface. This is because
the particle will have crossed the thermally definable boundary of the
solid and must be regarded as part of the liquid. For this simple
argument to be valid, it is essential, again, that the specific face
of the solid be {\em wetted} by the liquid, i.e. the face be actually
melting.  Since the typical energy of the surface vibrations is {\em
  exactly} $k_B T$, the particle's typical displacement free energy
cost will not exceed $k_B T$, hence $V^\ddagger = k_B T$. Further, we
define
\begin{equation}
  d_\sTS \equiv \frac{1}{2\sqrt{2}} l_\sTS, 
\end{equation}
where the numerical factor is chosen so that if the barrier is
parabolic at the top, then $d_\sTS = \la (\Delta x)^2 \ra^{1/2}$ in
the inverted potential at the saddle point. By definition, $m
(\omega^*)^2 (l_\sTS/2)^2/2 = k_B T$, where $\omega^*$ is the
under-barrier frequency, see Fig.\ref{rxn_sch}. After putting all this
together and recalling that $\tau_0 = a^2/6D = a^2 m/6 k_B T
\tau_\sauto = a^2/2 v^2_\sth \tau_\sauto$, one has by Eq.(\ref{t=t}):
\begin{equation} \label{dda2} \frac{d d_\sTS}{a^2} = \frac{1}{4
    \sqrt{6} \, \pi e} \simeq 0.01,
\end{equation}
universally. We therefore observe that it is possible to formulate a
purely {\em kinematic} criterion of melting, in terms of the ratio of
length scales characterizing molecular motions in the interface
region.  The numerical constants on the r.h.s.  of Eq.(\ref{dda2})
should not, perhaps, be taken too seriously; nevertheless the estimate
is expected to be accurate within a factor of 2 or so.

A number of comments are due here.  First of all, can one even apply a
transition state argument, when the barrier is so low? The answer is
yes, because of the high friction. Indeed, suppose for a moment there
were {\em no} barrier. Even so, the exit would be far from
instantaneous, being subject to (frequent) collisions with the nearby
molecules, just as are the molecular motions in the neighboring
liquid. Using the Smoluchowski diffusion limited reaction rate
expression (in 1D), one still gets the same basic scaling for the
escape rate: $k_\sescape \sim D/l^2_\sTS \sim v_\sth^2
\tau_\sauto/l^2_\sTS$ (see also \cite{LWF}). The overdamped character
of the molecular motion is essential in the present context, and so
one should like to estimate the actual value of the $l_\smfp/l_\sTS$
ratio, which is effectively the small parameter of theory. The
estimate in Eq.(\ref{t_t}) suggests that this ratio is indeed quite
small. Numerically it is of the order one hundredth, considering that
$d_\sTS/a \sim 0.1$. At any rate, molecular motions in liquids are
overdamped, near $T_m$, and so must be the motions in the
corresponding crystals, near $T_m$, because they are of comparable
(usually greater) density.

Despite simplifications due to the high friction, the shallowness of
the metastable potential complicates the interpretation of the simple
result in Eq.(\ref{dda2}). Since the transition state size is
virtually equal to the reaction path length, it probably makes little
sense to differentiate between $d_\sTS$ and $d$. For those same
reasons, the integral in Eq.(\ref{d}) is only meaningful in the {\em
  lowest} order in the reactant basin width, the expansion being
asymptotic of course. This leads to a simpler yet relation
\begin{equation} \label{ad} \we{d}_L \equiv \frac{d_L}{a} \simeq 0.1,
\end{equation}
where $d_L \simeq d \simeq d_\sTS$ stands for the amplitude of the
reversible motion in the molecular metastable minimum at the
liquid-solid interface. The subscript ``$L$'' alludes to
``Lindemann''; the parameter $d_L$ will be sometimes referred to as
the ``Lindemann length''.  Importantly, $d_L$ is a measure of the
displacement in the direction of the fastest escape, which is {\em
  perpendicular} to the surface. Further, $a$ signifies the molecular
spacing right at the interface. Strictly speaking this implies our
(volumetric) molecular spacing is a variable changing {\em
  continuously} across the interface. This appears reasonable as the
crystal and liquid density differ actually relatively little, not by
much more than the typical thermal density fluctuations in the
crystal, near the melting temperature. (See also the earlier mention
of the Oxtoby and Haymet's order parameter \cite{OxtobyHaymet}.)
Finally, the parameters $d$ and $d_\sTS$ should be dererminable in a
simulation, with a modest computational effort.

Are the simple relations (\ref{dda2}) and (\ref{ad}) consistent with
the general notions of surface melting? Yes, in a rather plain way. To
give an example, suppose the surface is rough, so that it has a
corner, or an edge. Clearly the vibrational amplitudes at corners and
edges are larger than those at extended flat faces, because there are
fewer neighbors. This is consistent with the expectation that corners
(and edges) melt first, i.e. at lower temperatures than say flat
faces. The relation in Eq.(\ref{ad}) is {\em quantitatively}
consistent with Valenta's calculations of the vibrational
displacements at the three distinct crystall faces of lead,
i.e. (110), (100), and (111) \cite{Valenta}.  These faces are known to
premelt at increasingly higher temperatures. According to
Ref.\cite{Valenta}, the vibrational displacements at {\em all} three
faces turn out ot be of nearly the same magnitude of $0.22 \AA$ or so,
at the respective premelting temperatures. In any event, one should
expect that denser packed, stiffer faces will exhibit lesser
vibrational displacements normal to the interface and therefore will
melt at increasingly higher temperatures. This qualitative notion is
consistent with the results of MD simulations on an FCC crystal, by
Kojima and Susa \cite{Kojima_Susa}, and on a BCC crystal, by Sorkin at
el. \cite{Sorkin}.

\subsection{A thermodynamic melting criterion in the bulk}
\label{L}

\begin{figure}[t]
\includegraphics[width=.95\figurewidth]{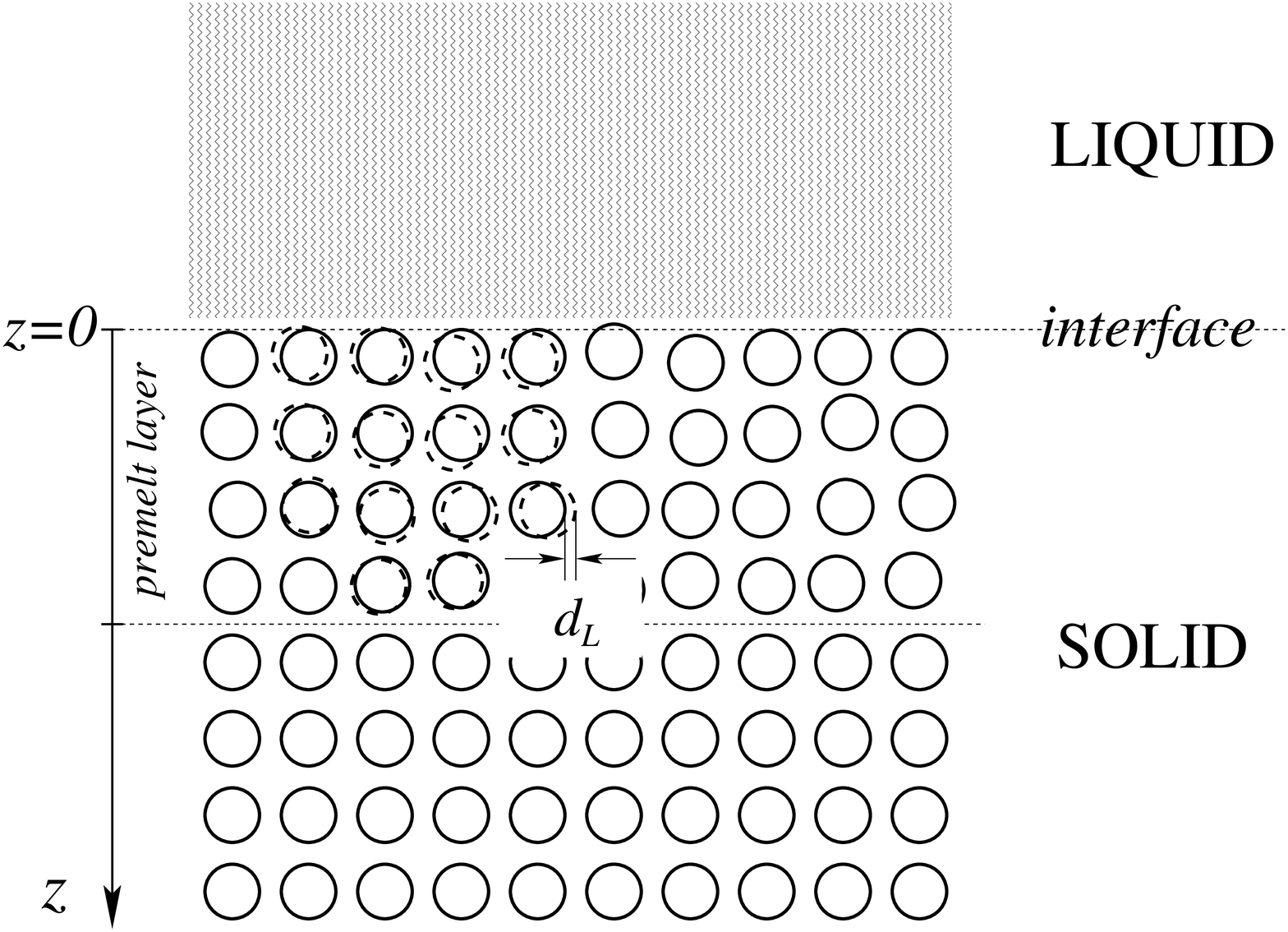}
\caption{\label{premelt} A cartoon of a solid-liquid interface is
  shown, including the premelt layer; the latter is characterized by
  some disorder. (Here, we ignore possible roughness or reconstruction
  of the surface.) The area containing the doubled circles illustrates
  a region undergoing a cooperative structural reconfiguration. The
  cooperativity is necessary to minimize density variations during the
  transition.  The solid and the dashed circles denote the atomic
  configurations before and after a structural transition. An
  individual, elemental displacement is of magnitude $d_L(z)$ mildly
  dependent on the distance to the interface. Each of these elemental
  displacements are nearly activationless; whereas the {\em total}
  cooperative event has a high barrier because the probability of a
  large number of {\em concerted} elemental events is low. }
\end{figure}

Let us now discuss the implications of the relations (\ref{dda2}) and
(\ref{ad}) for the molecular motion amplitudes in the {\em bulk} of
the solid. The typical {\em vibrational} displacements, $d_\svib(0)$,
at the surface and in the bulk, $d_\svib(z)$, will be certainly
comparable. The variable $z$ measures the distance from the interface,
see Fig.\ref{premelt}.  In the bulk, the greater lattice stiffness in
the bulk will be partially compensated by a smaller lattice constant,
save the substances that expand upon freezing. As a result, the local
$(d_\svib(z)/a(z))$ ratio, while also generically of about $0.1$ in
value in the bulk, is no longer expected to be strictly universal, in
contrast to that at the interface (see below). Now, one must bear in
mind that motions resulting in a locally different structure, may also
be present on the {\em solid} side of the interface. In the preceding
Subsection, we have computed the magnitude of the displacements $d_L$
such that would lead to the exit of an atom from the solid into the
liquid. In this Section, we will compute the magnitude of analogous
irreversible atomic displacements, but inside the solid. Here, the
atom also exits its present lattice site but to find itself in a {\em
  reconfigured} lattice, not in the liquid. Given the lattice has
reconfigured and is poised to accept a particle, the latter will
transfer in a nearly activationless fashion, similarly to exiting into
the liquid. The lattice reconfiguration itself is cooperative event
consisting of a large number of those elemental, nearly activationless
transitions occuring on the time scale $\tau_\sescape$. The latter
must occur in a concerted fashion, implying the cooperative
reconfigurations are rare and occur on much longer times scales:
\begin{equation}
  \label{t>>t}
\tau_\sstr \gg \tau_\sescape,
\end{equation}
where the index ``str'' indicates ``structural''.

First of all, do such rare structural reconfigurations take place in
the bulk?  We argue in the following that they do indeed, within
premelted layers. According to the surface calorimetric studies of
Santucci at el.  \cite{surf_calorimetry}, the excess entropy of the
``premelted'' surface layer of the Li (110) face, relative to the bulk
crystal, is about a {\em half} of the bulk liquid entropy excess, see
Fig.3(b) of Ref.\cite{surf_calorimetry}.  (The author is aware of
surface calorimetric estimates only on this particular substance,
however comparable values of the surface entropy are expected for
other materials as well, see \cite{surf_calorimetry} and references
therein.)  This clearly implies that the excess entropy, $s_c$, within
the premelt layer, is intermediate between that of the crystal and the
liquid. Furthermore, this entropy must decrease into the bulk, so as
to interpolate between the liquid and the solid values.  The latter is
zero. The relatively high density of states in the premelt,
corresponding to $s_c \sim k_B/2$ per particle, would be impossible to
account for by translations of vacancies: The vacancy concentration
would be too small, considering that the density and the
compressibility of the premelt are comparable to those in the solid
bulk. (The vacancy formation energies are in the eV range, of course,
see e.g. \cite{GrimvallSjodin}.) As a result, the room for molecular
translations is provided {\em not} by diffusing vacancies, but some
other structural degrees of freedom that involve more than one
particle. Another possibility is dislocations, whose significance
seems less straightforward to estimate than that of vacancies.  At the
present stage of theory, however, it appers that dislocations become
important closer to the point of {\em mechanical} melting, which they
most likely orchestrate in the first place.  The mechanical melting
temperature seems to be several tens of degrees above the bulk
thermodynamic melting point (see \cite{Dash_ContPhys, Dash_RMP,
  Sorkin_MSthesis} and references therein.)

What is the detailed microscopic nature of the configurational degrees
of freedom, in the surface premelted layer?  It is likely that they
correspond to transitions between long-living structures analogous to
the local metastable structures predicted by the random first order
transition (RFOT) theory of supercooled liquids and glasses, of
Wolynes and coworkers \cite{KTW, XW, LW_aging}. The transitions span
regions 3 to 6 molecular units across, and have been directly observed
by a variety of non-linear spectroscopies, see \cite{LW_RMP} for a
recent review.  The configurational entropy of a supercooled liquid,
just above the glass transition, is about $0.8 k_B$ per molecular
unit.  It is therefore comparable to that in the premelted layers. To
complete the analogy, note the viscosity of supercooled melts, near
$T_g$, is extremely high, consistent with the premelted layers
exhibiting progressively slower relaxation away from the interface
proper.

One may estimate, semi-quantitatively, how the magnitude of the
elemental displacements decreases away from the surface, as the
stiffness increases. The magnitude of the displacements will generally
depend on the distance from the interface, i.e.  $d_L = d_L(z)$, see
Fig.\ref{premelt}. Because a particle exits not into the liquid but
into a matrix poised to receive the particle, we no longer have the
convenience of the barrier being equal to the thermal energy. We may
say, nevertheless, that the barrier will increase into the bulk
because the lattice stiffness increases away from the surface. Let us
denote this $z$-dependent barrier as $ V^\ddagger(z)$, so that
\begin{equation}\label{V(0)}
  V^\ddagger(0) = k_B T_m.
\end{equation}

Now, it turns out that both $\tau_\sauto$ and $\tau_\sescape$ are
$z$-independent, because of the detailed balance: There is no net
particle transport at these time-scales. The actual net particle
transport occurs at the much slower time scale, $\tau_\sstr$, of the
extended structural transitions, and is consistent with the existence
of a density gradient in the premelt layer. (The detailed balance
argument above is only valid when the elemental and the structural
transitions are time scale separated, which is indeed true, in view of
Eq.(\ref{t>>t})). The constancy of $\tau_\sauto$ and $\tau_\sescape$,
together with Eqs.(\ref{tauescape}) and (\ref{V(0)}), yield that the
following quantity is invariant throughout the solid:
\begin{equation} \label{dV(z)} \left[\frac{d_L(z)}{d_L(0)} \right]^2
  e^{\frac{V^\ddagger(z)}{V^\ddagger(0)}} = e.
\end{equation}
This statement can be used to self-consistently determine the value of
the $d_L(z)/d_L(0)$ ratio, after one recalls that the activation
barrier $V^\ddagger$ arises from the elastic strain of the lattice:
Since the thermally relevant vibronic displacements are within about
one-tenth of the molecular spacing or less, the elastic energy, as a
function of the lattice strain, is dominated by the quadratic
component, i.e.:
\begin{equation} \label{V(z)} V^\ddagger(z) \simeq
  K(z)\left[\frac{d_L(z)}{a(z)}\right]^2 a^3(z) = K(z) a(z) d_L^2(z),
\end{equation}
Here, $K(z)$ is a local effective elastic modulus that depends both on
the isotropic compressibility and the shear modulus, and $a(z)$ is the
{\em local} volumetric spacing; in other words $V^\ddagger(z)$ gives
the elastic energy arising from the local strain $[d_L(z)/a(z)]^2$,
per molecular volume. (The ratio $[d_L(z)/a(z)]^2$ is a measure of the
thermally averaged square of the dimensionless elastic strain used
here as in the standard elasticity theory \cite{LLelast}.)  One
obtains, as a result:
\begin{equation} \label{d(z)} \left[\frac{d_L(z)}{d_L(0)} \right]^2
  e^{\zeta(z) \left[\frac{d_L(z)}{d_L(0)}\right]^2} = e,
\end{equation}
where, again, $[d_L(0)] \simeq 0.1 a$ and the parameter $\zeta(z)$
gives the stiffness of the lattice, at distance $z$ from the
interface, relative to its value at the interface:
\begin{equation} \label{z(z)} \zeta(z) = \frac{K(z)a(z)}{K(0)a(0)}.
\end{equation}
Note that $K(0) a(0) d_L^2(0) = V^\ddagger(0) = k_B T_m$. We stress
again that the (very low) barriers, implied in Eqs.(\ref{V(z)}) and
(\ref{d(z)}), correspond to elemental translations that would occur
{\em given} another structural state exists and therefore apply to
{\em all} of the bulk. Because the elemental events are subject to the
existence of an underlying structural transition in the corresponding
region, it is appropriate to term the displacements $d_L$ as {\em
  fiduciary} displacements, or {\em presumable} displacements.

\begin{figure}[t]
\includegraphics[width=.95\figurewidth]{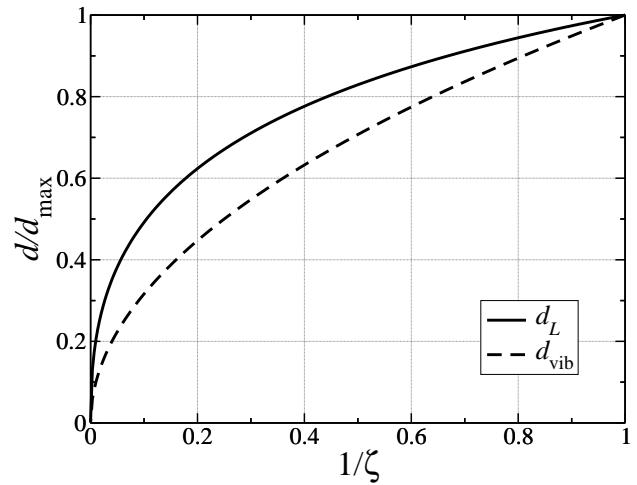}
\caption{\label{zda} Shown are the dependences of the Lindemann and
  the vibronic displacements, relative to their maximal value
  $d_\smax$, achieved at the surface, as functions of the
  dimensionless stiffness $\zeta$. The latter is the product of the
  elastic constant and the lattice constant, in terms of the minimum
  value of this product achieved at the surface.}
\end{figure} 
It is easy to solve, numerically, Eq.(\ref{d(z)}) for the
$d_L(z)/d_L(0)$ ratio as a function of the dimensionless stiffness
$\zeta(z)$, with the result shown in Fig.\ref{zda}.  The fiduciary
displacement $d_L(z)$ should be compared to the thermal vibrational
amplitude $d_\svib(z)$, which is fixed by the equipartition theorem:
\begin{equation} \label{d_vib} K(z) a(z) d^2_\svib(z) = k_B T.
\end{equation}
The l.h.s. of the first equation gives the usual elastic energy
density of the solid, times the volume of a unit cell.  As follows
from the discussion in the preceding Subsection,
\begin{equation}
  d_\svib(0) = d_L(0) \equiv d_\smax.    
\end{equation}
The vibronic displacement $d_\svib(z)$ depends on the stiffness in a
very simple way, and is shown in Fig.\ref{zda}, alongside the
$\zeta$-dependence of the fiduciary, Lindemann diplacement. Note that
here we did not have to treat explicitly the anisotropy of elastic
properties, as long as we were consistent in measuring the
displacements $d_L(0)$ and $d_\svib(0)$ {\em normal} to the free
surface.

Since the $z$-variable is a dummy label, it is appropriate to simply
think of the displacements $d_L$ and $d_\svib$, relative to their
maximal values achieved at the surface, as functions of the
dimensionless stiffness $\zeta$. According to Fig.\ref{zda}, it is
always true that
\begin{equation} \label{d<d} d_\svib(z) < d_L(z), \hspace{5mm} (z > 0),
\end{equation}
c.f. Eq.(\ref{t>t}). We therefore observe that solids may be defined
as collections of molecules in which the magnitude of purely
vibrational excitations is always smaller than the extent of the
fiduciary structural displacements. As a result, a solid is
(typically) capable of sustaining vibrational excitations without
irreversible structural changes.  According to Fig.\ref{zda}, a
stabler solid corresponds to greater values of the stiffness parameter
$\zeta$, corresponding to larger values of the difference $(d_L -
d_\svib)$.  One may therefore interpret Eq.(\ref{d<d}) as saying that
the stiffness of a lattice that allows for alternative structures, is
somewhat smaller than the stiffness of the corresponding lattice which
is strictly elastic, i.e. mechanically stable. We thus observe that
vanishing of the shear modulus is {\em not required} in order for the
lattice to be locally mechanically unstable. Now, the limiting
situation where $d_\svib(z) = d_L(z)$, at $z=0$, bears a dual meaning:
(a) it signifies melting and (b) it defines the boundary between the
solid and the liquid phase.  In view of Eq.(\ref{dV(z)}), the
statements in this paragraph are valid regardless of the elastic free
energy being strictly quadratic in the displacement. Finally note that
Eqs.(\ref{d(z)}) and (\ref{d_vib}) are laws of corresponding states
for the structural and vibrational displacements in a solid.

We have thus found that formulating a criterion of melting and
defining a solid, in the first place, requires introducing {\em
  fiduciary} structural modes, in the sense that these modes may or
may not be readily observed in a real material. We have argued that
such modes probably exist in a sufficiently extended premelt
layer. Otherwise, there is still a way to force the solid to sample
the spectrum of those structural excitations, even at temperatures
below the melting point, namely by quenching the corresponding
substance from its liquid state into a frozen glass. Many polymers
simply do not form crystals, and are difficult to characterize
morphologically in a definite way, in the first place. In these, there
is no shortage of structural transitions. A similar comment applies to
folded proteins.  Certain substances, that are very difficult to
supercool, such as ice, {\em amorphize} under pressure (see
e.g. \cite{Strassle}). Finally, energetic particle beams may be
employed to locally force a crystal into a structurally excited
state. The possibility of such externally induced structurally excited
states is actually a physical realization of a Maxwell construction!
To give an analogy, the Maxwell construction in a non-ideal gas also
uses presumable states, i.e. such that $(\prtl p/\prtl V)_T > 0$, in
deriving a criterion of a first order transition,
i.e. boiling/condensation.

Now, suppose there {\em is} a significant premelt layer on the
surface. Direct spectroscopic observations of these are intrinsically
difficult, since a beam (neutron or X-ray) sufficiently intense to
produce useful signal, would also inevitably heat up and liquefy the
surface. On the other hand, surface, near-atomic resolution
techniques, such as those employed by Israeloff
\cite{RusselIsraeloff}, may be helpful in characterizing structural
surface modes. Importantly, a conclusive surface study must be
non-linear so as to sense {\em dynamic} heterogeneity, in order to
distinguish irreversible structural relaxations from purely
vibrational excitations. 

Finally, for completeness, one should like to see whether the
arguments above would also be robust in systems other than
single-component liquids.  Unfortunately, mixtures necessitate
considering many additional factors, well beyond the scope of this
article, primarily owing to the details of interaction between the
constituents of the mixture, phase separation etc.  Here, we will
limit ourselves to a few remarks.  Suppose the molecules are
(chemically identical) rigid, relatively weakly interacting rods.
Clearly the melting will occur from the face parallel to the
orientation of the rods, otherwise the molecule would have to slide
out of the matrix, which would be too costly energetically. (We assume
the rods are closely packed, in a colinear fashion.)  The ``melting
displacement'' $d$ is therefore perpendicular to the rods'
orientation, and, consequently, is also about one tenth of the {\em
  monomer} size, but probably not universally.  Analogous logic
applies to a crystal made of weakly interacting disk-shaped molecules,
which will melt off the crystallographic plane parallel to the disk
planes. (This is in the case when the discs are stacked in a coplanar
manner.)  The case of mono-valent ionic melts is probably adequately
viewed as that of a single-component substance with a slightly
elongated molecule, since the melt will tend to be locally
electrically neutral, on average. Again, we recover the basic estimate
from Eq.(\ref{ad}) where $a$ is the volumetric spacing between the
distinct moieties, {\em not} the unit cell size; $d$ is the
displacement between the neighboring atoms.

\section{Activated motions in supercooled liquids as local melting}

The analysis of Subsection \ref{L} was conducted on the assumption
that solids exist, of course. In other words, the argument by itself
could not be used to establish the existence of a solid state, but
could only give the conditions of stability of a solid once it is
formed. The details of solidification, below $T_m$, are known to
depend on how fast one cools the substance and on the liquid's
viscosity. If either of the cooling speed or the viscosity is high
enough, the nucleation of the crystalline phase will be prevented,
with the liquid finding itself in a supercooled state for a
significant amount of time.

Here, we will estimate the temperature dependence of the elemental
displacements in a liquid that was cooled below its fusion point, but
has failed to crystallize. Since the {\em structural equilibration has
  never occured}, displacements {\em must} take place such that they
imply a change of the local structural state.  Since these
displacements are at the mechanical stability edge of the material, we
may also denote their magnitude with $d_L(T)$, where we explicitly
indicate the temperature dependence of the magnitude of the elemental
displacements. (Clearly, $d_L(T_m) = d_L(z=0)$.)  Therefore in a
supercooled liquid, the displacements $d_L$ are no longer fiduciary,
but strictly factual. On the other hand, the regular liquid state,
that does not discriminate between molecular displacements of less
than size $a$ (recall our ``labelling'' discussion), now becomes a
fiduciary state!  (For reference, we point out such a regular liquid
state explicitly arises, for instance, in the DFT study of aperiodic
crystals in Ref.\cite{dens_F1}, or of the mean-field Potts glasses
\cite{MCT1}.) We may use this new fiduciary state to write the law of
corresponding states from Eq.(\ref{d(z)}) at a temperature $T$ {\em
  below} $T_m$:
\begin{equation} \label{d(z)T} \left[\frac{d_L(z, T)}{d_L(0, T)}
  \right]^2 e^{\zeta(z, T) \left[\frac{d_L(z, T)}{d_L(0, T)}\right]^2}
  = e.
\end{equation}
Here, even though the $z$-variable is dummy, it is strictly implied
that $z = \infty$, i.e. the interface is infinitely remote from the
bulk, by construction. Since the universal quantity $\tilde{d}_L
\equiv d_L(0)/a(0)$ is, by its very meaning, temperature independent,
it is convenient to rewrite the above relation in the following form:
\begin{equation} \label{dz_corr} \frac{(d_L/a)^2}{\tilde{d}_L^2}
  e^{\zeta(K a^3/T) \frac{(d_L/a)^2}{\tilde{d}_L^2}} = e,
\end{equation}
where
\begin{equation} \label{z(z)corr} \zeta(K a^3/T) \equiv \frac{K(T)
    a^3(T)/k_B T}{K(T_0) a^3(T_0)/k_B T_0}.
\end{equation}
As a reminder, $K(T_0) a^3(T_0) \tilde{d}^2/k_B T_0 = 1$, for any
$T_0$. While, in principle, the interface at any temperature $T_0 \le
T_m$ may be used as a reference state, in practice one can measure the
stiffness only at the physical melting temperature $T_m$; therefore
most conveniently one would set $T_0 = T_m$ in Eq.(\ref{dz_corr}).
Again, we have arrived at a law of corresponding states, even though
the variables are distinct from those entering Eq.(\ref{d(z)}): The
displacements are in relation to the lattice spacing; the running
coupling constant $K a^3/T$, governing the displacements magnitude, is
dimensionless too, and is of the order 1. The latter suggests that
below the melting temperature (but above vitrification!) the system
remains at a delicate balance between energetic and entropic
contributions to its free energy (see also \cite{XW} and the
discussion of the surface energy below).

As already mentioned, the RFOT theory has built a constructive,
microscopic picture of structural relaxations in deeply supercoooled
liquids and frozen glasses. We learn from the RFOT theory that one may
think of structural rearrangmeents in deeply supercooled liquids as
activated growth of distinct aperiodic phases within each other.  The
corresponding activation profile is \cite{KTW, XW}:
\begin{equation} \label{F(N)eq} F(N)|_{T > T_g} = \gamma \sqrt{N} - T
  s_c N,
\end{equation}
where $N$ is the number of particles in the nucleus of the new
structural state within the previous structural state.  The
configurational entropy (the $-T s_c N$ term) drives the transitions,
while the barrier arises due to the mismatch energy penalty, $\gamma
\sqrt{N}$, between distinct states. This mismatch may be thought of as
the surface tension of the {\em domain wall} separating the two
alternative structural states.  As already mentioned, the elemental
displacements occur on a conditional basis, if an alternative
structural state is present locally.  As a result, the full RFOT
structural relaxation rate is:
\begin{equation} \label{tRFOT}
  \tau_\sRFOT^{-1} = \tau^{-1}_\sescape(T, z=\infty) e^{-F^\ddagger_\tRFOT/T}.
\end{equation}
(The time scale $\tau_\sstr$, introduced earlier, was an analog of the
time scale $\tau_\sRFOT$ but in the context of structural relaxations
in premelted crystalline layers.)

The elemental translations, corresponding to the length $d_L$, play an
important role in the RFOT theory, for a number of reasons. First of
all, the corresponding length scale arises as an order parameter
during a first order cross-over from the ``regular'' liquid state to
the one where metastable structures persist for a discernible
time. (For a deep enough quench, this time is given by
$\tau_\sRFOT^{-1}$ from Eq.(\ref{tRFOT}).) In the course of this
cross-over, the localization length of a molecule jumps from infinity
to a finite value, i.e.  $d_L$ itself \cite{dens_F1}. This
localization length depends only weakly on the temperature/density, as
the DFT study in Ref.\cite{dens_F1} suggested. The near constancy of
the Lindemann ratio $d_L/a$ turned out instrumental in establishing
the near universality of the surface tension coefficient between
locally competing liquid structures in deeply supercooled
liquids. This surface tension was computed by Xia and Wolynes (XW)
\cite{XW} without adjustable parameters: it depends only
logarithmically on the Lindemann ratio leading to the following
expression for the tension coefficient $\gamma$:
\begin{equation} \label{gamma} \gamma = \frac{2\sqrt{3\pi}}{2} k_B T
  \ln\left[\frac{(a/d_L)^2}{\pi e}\right].
\end{equation}
(The notations are from Refs.\cite{LW, LW_RMP}.)  This result, among
other things, enabled XW to calculate the numerical value of the
barrier for structural reconfiguration from Eq.(\ref{F(N)eq}) leading
to specific estimates of the size of a cooperatively rearranging
region. This size grows from a few molecular units (``beads''), near
$T_m$, to roughly 200 beads near the glass transition, implying each
region is about 5-6 beads across near $T_g$. (A bead typically
consists of a few atoms; for a detailed discussion see \cite{LW_soft},
and also \cite{LSWdipole}).

The present arguments enable one to compute explicitly the temperature
dependence of the Lindemann ratio, based on the relation in
Eq.(\ref{dz_corr}). The result is shown, again, in Fig.\ref{zda},
where $\zeta = k_B T/(K(T) a^3(T))$ now, see Eq.(\ref{z(z)corr}). The
lattice spacing of the liquid will decrease, and the elastic constants
will increase, with lowering the temperature, albeit weakly.  Here of
course, we mean the high frequency elastic constants, which are
definable on time scales shorter than the life-times of the
long-living metastable structures. For most substances, the ratio
$T_g/T_m \sim 2/3$, empirically. As a result, the parameter
$\zeta^{-1}$, in Fig.\ref{zda}, will decrease at most to the value of
0.67, before the liquid freezes into a glass, leading to
$d_L(T_g)/a(T_g) \simeq 0.9$, according to the figure.  We have thus
established, on a first principles basis, that the $d_L/a$ ratio
indeed varies little with temperature, at most by 10\%. This
corroborates the use of the Lindemann criterion, by the RFOT theory,
to establish the near universality of the surface tension on the basis
of a near universality of the molecular displacement at the mechanical
stability edge, relative to the molecular spacing. Finally, the
computed temperature dependence of the Lindemann length $d_L$ should
be measurable by neutron scattering, as in Ref.\cite{MezeiRussina}.

\begin{figure}[t]
\vspace{4mm}
\includegraphics[width=.90\figurewidth]{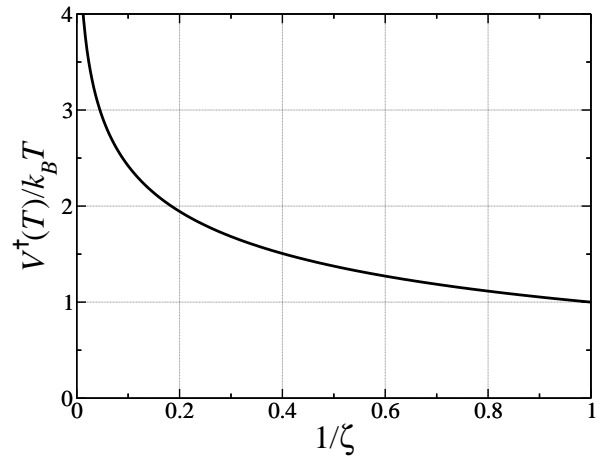}
\caption{\label{V_T} Shown is the ratio of the energy barrier to the
  temperature, for an elemental translation, as a function of
  $\zeta$. }
\end{figure}

Further we can see explicitly what parameters in the problem drive the
surface energy $\gamma$, by substituting the running value of the
$d_L/a$ ratio from Eq.(\ref{dz_corr}) into the logarithm in
Eq.(\ref{gamma}):
\begin{equation}
  k_B T \ln\left[\frac{(a/d_L)^2}{\pi e}\right] 
  =  k_B T [\ln(4\sqrt{6}) -1] + V^\ddagger(T),
\end{equation}
where $V^\ddagger(T) = K(T) a(T) d_L^2(T)$ is the actual energy
barrier for a typical elemental translation, as stemming from the
lattice strain; also, $[\ln(4\sqrt{6}) -1] \simeq 1.28$.  It follows
from the equation above that the energetic and entropic contributions
to the surface tension are actually comparable, at all temperatures
above vitrification: $T > T_g$. For the reference, we show in
Fig.\ref{V_T} the $V^\ddagger(T)/k_B T$ ratio, which is easily
computed from Eq.(\ref{dz_corr}), as a function of $\zeta$. At
$\zeta^{-1} \simeq .67$, $V^\ddagger \simeq 1.21 k_B T$. It is not
entirely clear, at present, what significance should be attributed to
the $\zeta^{-1}$-interval that corresponds to temperatures {\em below}
vitrification.

Finally, it seems instructive to make an explicit estimate of the
$\tau_\sescape^{-1} (T, z=\infty)$.  As already mentioned, the number
of collisions per unit time, for a harmonically confined particle,
scales linearly with energy, leading to
$\tau_\sauto(T)/\tau_\sauto(T_m) = T_m/T$. This and
Eqs.(\ref{tauescape}-\ref{dda2}) yield:
\begin{equation}
  \tau_\sescape^{-1} = \sqrt{3/8} \, \frac{D (T_m)}{\pi d d_\sTS} e^{-V^\ddagger/k_B T} 
  =  \tau_0^{-1}(T_m) e^{1-V^\ddagger/k_B T},
\end{equation}
where $D(T_m)$ and $\tau_0(T_m)$ are the regular diffusion constant
and the equilibration time scale $\tau_0$, of the corresponding liquid
at the melting temperature $T_m$.  $\tau_0(T_m)$ is about ten
picoseconds, and the exponential part actually does not lead to an
activated temperature dependence, but a much weaker one, in view of
Fig.\ref{V_T}.

\section{Summary}


We have established the existence of a universal criterion of melting,
in terms of the ratio of length scales characterizing the escape of a
particle from its current metastable configuration. The criterion is
therefore purely kinematic. The obtained quantitative results are
consistent with earlier studies of displacements at crystal surfaces,
for several specific substances \cite{Valenta, Kojima_Susa, Sorkin}.
The said length scales are closely related to the vibrational
amplitudes in the crystal bulk, which was argued to underlie the
otherwise puzzling consistency of the empirical Lindemann criterion.

A proper treatment of bulk mechanical stability has been performed,
and has required considering ``fiduciary'' alternative structural
states in the lattice. Such alternative states are known to exist in
supercooled liquids and glasses, and were argued here to exist in a
premelted layer at the liquid-crystal interface. A proper criterion of
mechanical stability was formulated; it stipulates that the
vibrational molecular displacements be less than the elemental
displacements that would occur during the multi-particle structural
transitions. We have seen that vanishing of the shear modulus is not
necessary for the lattice to be mechanically unstable, consistent with
the apparent high configurational entropy of premelt layers. Direct
observation of cooperative rearrangements in such layers is difficult,
but may be possible with available surface techniques.

We have computed the dependence of the elemental displacements on the
material's stiffness, and on the temperature. When taking place in
supercooled liquids, these can be measured, for instance, by neutron
scattering.

{\em Acknowledgments:} The author thanks Peter G. Wolynes for
stimulating discussions and useful suggestions. This work has been
funded in part by the GEAR Program and the New Faculty Grant at the
University of Houston.

\vspace{-3mm}

\bibliography{/Users/vas/Documents/tex/ACP/lowT}

\vspace{-3mm}

\end{document}